\documentclass[a4paper, 10pt, twocolumn]{article}

\usepackage[utf8]{inputenc}         
\usepackage[]{babel}          

\usepackage{graphicx}               
\usepackage{tabularx}               
\usepackage{multirow}               
\usepackage{url}                
\makeatletter
\newcommand\footnoteref[1]{\protected@xdef\@thefnmark{\ref{#1}}\@footnotemark}
\makeatother

\usepackage[table]{xcolor}
\usepackage{colortbl}
\definecolor{headercolor}{RGB}{70,130,180}
\definecolor{rowcolor}{RGB}{220,220,220}

\usepackage[small,bf]{caption2}     
\usepackage{parskip}
\usepackage{titlesec}
\usepackage{amsmath}                

\titleformat{\section}{\normalfont\large\bfseries}{\thesection}{}{}
\titleformat{\subsection}{\normalfont\large\bfseries}{\thesection}{}{}
\titleformat{\paragraph}{\normalfont\bfseries}{\theparagraph}{}{}
\titlespacing{\section}{0pt}{6pt}{-1pt}
\titlespacing{\subsection}{0pt}{3pt}{-1pt}
\titlespacing{\paragraph}{0pt}{3pt}{-1pt}

\newcolumntype{Y}{>{\centering\arraybackslash}X}    

\begin{document}

\date{}                                         

\title{\vspace{-8mm}\textbf{\large
Joint sentiment analysis of lyrics and audio in music }}

\author{
Lea Schaab$^1$, Anna Kruspe$^2$\\
$^1$ \emph{\small Technische Hochschule Nürnberg Georg Simon Ohm, Email: schaable79981@th-nuernberg.de
}\\
$^2$ \emph{\small Munich University of Applied Sciences,
Email: anna.kruspe@hm.edu } } \maketitle

\thispagestyle{empty}           
\section*{Motivation}
\label{sec:motivation} 

Music is often described as the language of emotions, and numerous studies have confirmed that listeners perceive music as an expression of feelings \cite{juslin2001psychological}. The expression of emotions and mood regulation, alongside personal self-discovery and social connectedness, are among the primary reasons people listen to music \cite{schaefer2013psychological}. The ``Engaging with Music 2022'' report by the International Federation of the Phonographic Industry revealed that 69\% of respondents consider music important for their mental health \cite{ifpi2022engaging}.


With the rapid growth in the digital availability of music, the relevance of automated music information retrieval systems that allow searching and organizing music based on specific factors has increased. While traditional approaches to managing music rely on metadata such as song titles, artist or album names, they often have limited applicability for most music-related queries \cite{huron2000perceptual}. Retrieving music based on perceived emotions has gained significance in recent years, though it is still in its early stages of development. This opens up a variety of potential applications, including music recommendation systems, music search, music visualization, automatic music generation, music therapy, or selecting suitable background music for various contexts such as restaurants, advertising, and film production \cite{huron2000perceptual, han2022survey}.

For these reasons, Music Emotion Recognition (MER) is a popular research topic within the Music Information Retrieval (MIR) community. Commonly, emotions are extracted solely on the basis of the auditory content of songs. In contrast, textual lyrics are an under-researched modality in many MIR tasks \cite{humphrey,Kruspe2014KeywordSI, Kruspe2016RetrievalOT}. However, lyrics play an important role for eliciting emotional responses in listeners \cite{stratton1994affective, juslin2004expression}. In this paper, we will therefore look into emotion recognition both on the auditory level as well as the corresponding textual lyrics. We will use existing machine learning models for the two separate modalities, compare results, and will then analyze options for fusing those.


\begin{table}[htbp]
    \centering

    \caption{MIREX Mood clusters \cite{hu2008mirex}}
    \vspace{2mm}
    \label{tab:mood_clusters}
    \begin{tabular}{|c|p{6.5cm}|}
        \hline
        \textbf{Cluster} & \textbf{Mood} \\
        \hline
        Cluster 1 & passionate, rousing, confident, boisterous, rowdy \\
        \hline
        Cluster 2 & rollicking, cheerful, fun, sweet, amiable/good-natured \\
        \hline
        Cluster 3 & literate, poignant, wistful, bittersweet, autumnal, brooding \\
        \hline
        Cluster 4 & humorous, silly, campy, quirky, whimsical, witty, wry \\
        \hline
        Cluster 5 & aggressive, fiery, tense/anxious, intense, volatile, visceral \\
        \hline
    \end{tabular}

\end{table}

\section*{Theoretical basics}
\label{sec:theory}
Emotions are a key research topic in psychology, leading to the development of two main approaches for classifying emotions in music sentiment analysis: the categorical and dimensional approaches.

The categorical approach classifies emotions into distinct categories based on basic emotions theory, suggesting a limited number of primary emotions like happiness, sadness, anger, fear, and disgust, from which all other emotions derive \cite{oatley1987towards}. This approach has been applied in studies and frameworks such as the Music Information Retrieval Exchange (MIREX) Audio Mood Classification task, which categorizes emotions into five clusters with related adjectives \cite{hu2008mirex} as shown in table \ref{tab:mood_clusters}. However, its drawbacks include limited resolution compared to the spectrum of music emotions perceived by humans and susceptibility to ambiguities due to linguistic descriptions \cite{chen2011music}.

\begin{figure}[hbt]
    \begin{center}
        \includegraphics[width=4.5cm]{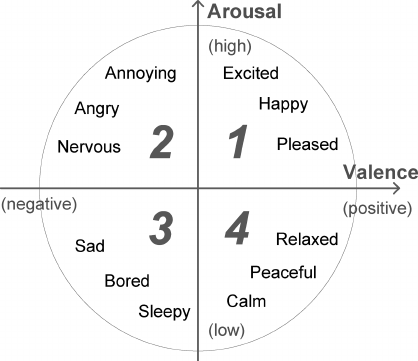}
    \end{center}
    \caption{The 2D valence-arousal emotion space \cite{yang2012machine}}
    \label{fig:russell}
\end{figure}

On the other hand, the dimensional approach views emotions as points within a multidimensional space, primarily using Russell's circumplex model of affect \cite{russell1980circumplex}. This model utilizes two dimensions: valence (pleasantness) and arousal (intensity) to map emotions, offering a simple yet effective way to organize and compare different emotions, as illustrated in figure \ref{fig:russell}. Despite its advantages, the dimensional approach might blur important psychological distinctions and might not fully capture all emotions with just a few dimensions \cite{chen2011music}. Some researchers suggest adding a third dimension, like dominance, to provide a more comprehensive emotional representation, though this adds complexity and is not universally accepted.

\section*{Data}
\label{sec:data}
The selection of data sets was based on criteria such as the simultaneous availability of audio and text, size of the data set, annotation quality, and compatibility with established emotional models. Two data sets from the MOODetector project \cite{cisucMOODetector} were chosen that best met these criteria.

\paragraph{VA Data Set} This data set, first presented in 2016 in \cite{malheiro2016bi}, includes 133 audio excerpts and their corresponding lyrics annotated according to Russell's circumplex model, i.e. valence and arousal values. Annotators were asked to label lyrics and music individually.

\paragraph{MIREX-like Data Set} This data set from 2013 contains 903 audio clips, 764 lyrics, and 193 MIDI files, with metadata organized in a CSV file and annotations based on the All Music Guide \cite{panda2013multi}. 

\paragraph{Data Preprocessing} 

To standardize both data sets, the VA data set's textual annotations were aligned to the Valence-Arousal quadrants based on their labeled values. For the MIREX-like data set, the annotations were categorized into the corresponding quadrants, utilizing valence and arousal values derived from the ANEW lexicon \cite{bradley1999affective}. The mappings can be found under \url{https://github.com/annakaa/sentiment_mappings}. Other preprocessing steps included augmenting the data set with lyrics, publication years and genres, normalizing text and audio, and correcting inconsistencies and duplicates.



\section*{Audio model}
\label{sec:audiomodel}
For the audio modality, we employed a model by USC SAIL which performed best in the ``Emotions and Themes in Music'' MediaEval task in 2020, which involves predicting tags from audio to identify music-related moods and themes  \cite{bogdanov2020mediaeval}. The USC SAIL model employs a Short-Chunk CNN with residual connections and was trained on datasets including MTG-Jamendo and MagnaTagATune \cite{uscsail}.

To adapt the MediaEval tags to the four quadrants of Russell's circumplex model, as organized in the datasets, the procedure described previously for the MIREX dataset was followed. Mood-related tags were assigned to the quadrants based on valence and arousal dimensions according to the ANEW lexicon. The mappings are also available under \url{https://github.com/annakaa/sentiment_mappings}.

During the adaptation process, it was noted that the tags ``love'' and ``sexy'' frequently appeared across all quadrants, leading to their exclusion from further consideration and reassignment to theme tags. Tags ``epic'' and ``heavy'' were predominantly associated with Quadrant Q2, while ``meditative,'' not present in the ANEW lexicon, was subjectively assigned to Q4 based on personal judgment.  

\section*{Text models}
\label{sec:textmodels}
Four models from the Hugging Face platform were evaluated for sentiment analysis of song lyrics:

    \textbf{finetuning-sentiment-model-4500-lyrics} This model is specifically designed for sentiment prediction in song lyrics, based on a fine-tuned distilBERT model. It classifies lyrics as Negative or Positive. \cite{finetuning-sentiment-model-4500-lyrics}\\
    \textbf{sentiment-roberta-large-english (SiEBERT)} A fine-tuned version of RoBERTa-large, optimized for performance. SiEBERT delivers binary predictions and was trained and evaluated across 15 datasets. \cite{sentiment-roberta-large-english}\\
    \textbf{bert-base-uncased-poems-sentiment} Aimed at classifying poems into categories (negative, positive, neutral, mixed), this model's adaptability to the poetic language of song lyrics was explored. \cite{bert-base-uncased-poems-sentiment}\\
    \textbf{bert-base-uncased\_finetuned\_sentiments} Trained to classify texts into six different emotions (anger, fear, joy, love, sadness, surprise), offering a multi-emotion perspective compared to binary sentiment models. \cite{bert-base-uncased_finetuned_sentiments}

A significant challenge was the maximum token count limitation of 512 tokens, typical for BERT models. To manage lyrics exceeding this limit, texts were segmented into chunks, processed separately, and then aggregated for an overall sentiment prediction. This chunking process allowed for sentiment analysis on longer texts, thus expanding the scope of analysis.

\section*{Experimental results}
\label{sec:experiments}
\begin{figure}
    \centering
    \includegraphics[width=1\linewidth]{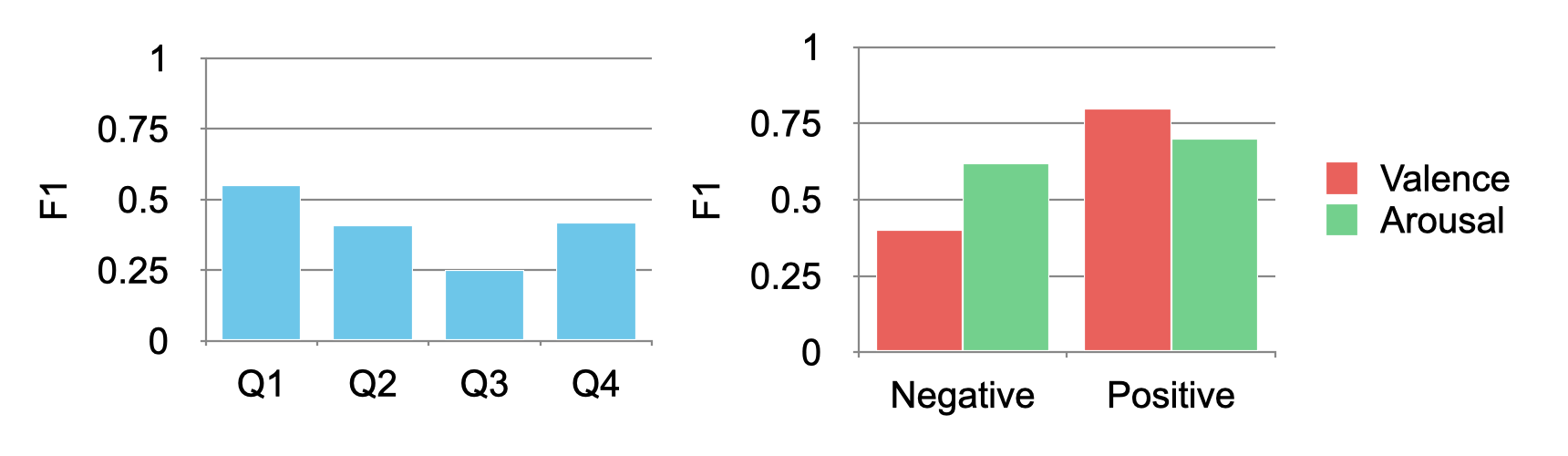}
    \caption{Results of the audio-only model. Left: Quadrants of the circumplex model, right: Binary results for valence and arousal.}
    \label{fig:audio_results}
\end{figure}

\subsection*{Audio only}
The USC SAIL model's performance in audio sentiment analysis was evaluated across Russell's circumplex model quadrants, and the valence and arousal dimensions. Results are shown in figure \ref{fig:audio_results}.

For the classification into Russell's circumplex model quadrants, Quadrant Q1 showed the best results while Quadrant Q3 had the lowest performance metrics.
In the valence dimension, the model demonstrated a clear strength in identifying positive sentiments, with average metrics across positive and negative sentiments around 60\%.
The model's performance in the arousal dimension averaged higher than in the valence dimension. Arousal may be easier to determine in the audio modality due to a clearer musical expression of calm vs. excitement.

\begin{figure}
    \centering
    \includegraphics[width=1\linewidth]{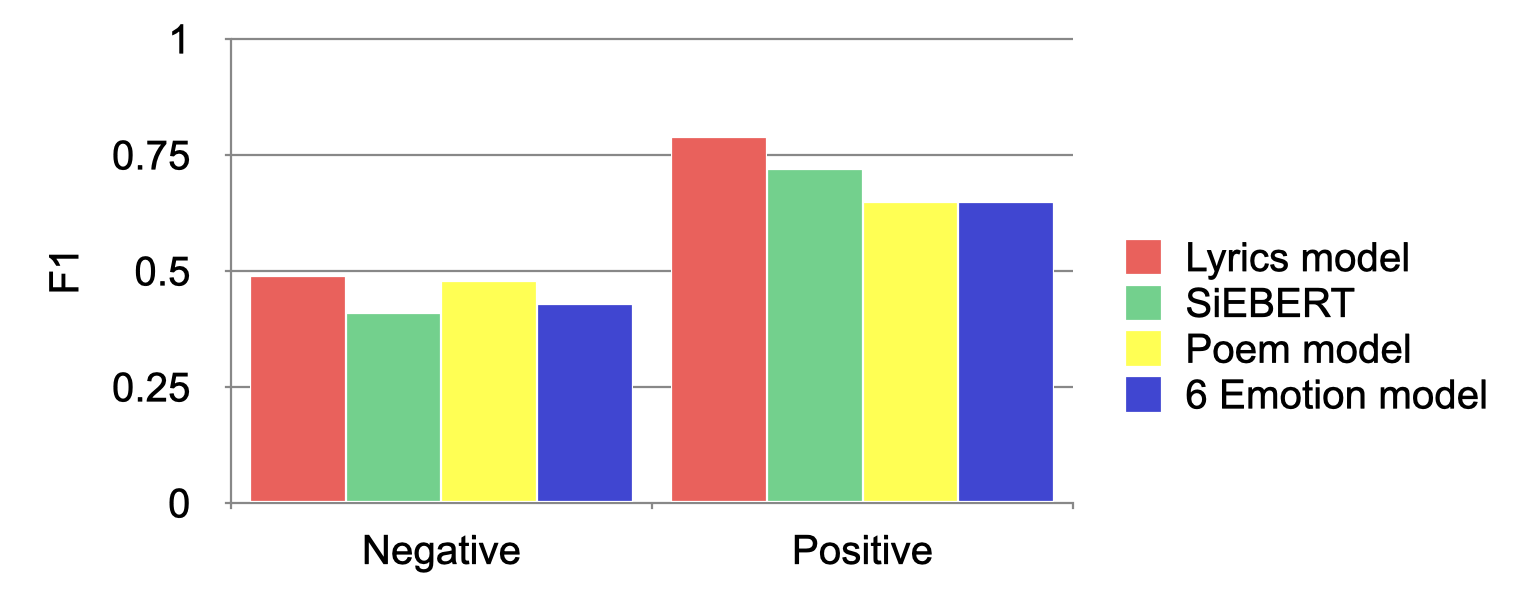}
    \caption{Results of the various text-only models.}
    \label{fig:text_results}
\end{figure}

\subsection*{Text only}
The four models were assessed for sentiment analysis in song lyrics, focusing solely on valence due to consistent availability across models. The results are shown in figure \ref{fig:text_results}. The Lyrics Model excelled in detecting positive sentiments, showing higher precision and F-Score for positive over negative emotions, indicating strong positive sentiment identification. SiEBERT, with a broader training base, slightly lagged behind the Lyrics Model in both emotion types, suggesting specialized models like the Lyrics Model perform better with specific tasks like song lyrics.

The Poem Model, designed for poetic content, showed high precision in positive emotion classification, even outperforming the Lyrics Model in some aspects, underscoring its suitability for lyric analysis due to their poetic nature. The 6-Emotions Model, covering a wider emotional range, displayed a mixed performance with better recall for negative emotions but lower precision, highlighting its capability to identify negative sentiments but occasionally misclassifying positive ones as negative.

In conclusion, the Lyrics and Poem Models delivered strong results, with the latter possibly more adept at negative emotion detection. The 6-Emotions Model showed varied precision and recall, indicating a nuanced approach to sentiment classification in lyrics. Surprisingly, the best lyrics model surpasses the results of the audio model, confirming the relevance of lyrics for the valence recognition task.

\begin{figure}
    \centering
    \includegraphics[width=1\linewidth]{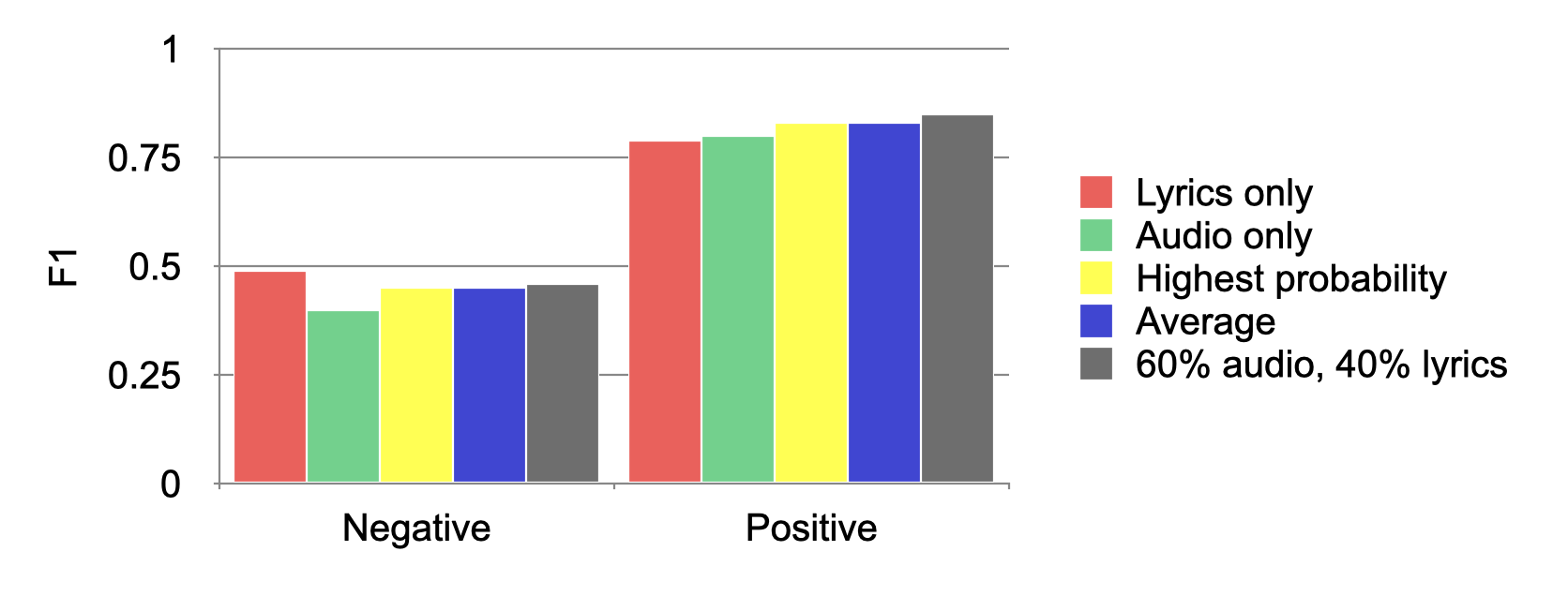}
    \caption{Results of different model fusion strategies.}
    \label{fig:fusion_results}
\end{figure}

\subsection*{Fusion of the two modalities}
Three different methods for fusing audio and text results were investigated to enhance the performance of music sentiment analysis systems by leveraging both modalities. These methods included: class selection based on the highest probability, averaging predictions, and weighted combination. Results are shown in figure \ref{fig:fusion_results}.

The first method utilized the predictions from both the audio and text models for negative and positive classes, selecting the class with the highest probability. This approach found that in about 19\% of cases, audio classifications were chosen due to higher probability, while text classifications were preferred in approximately 81\% of cases.

The second method involved averaging the predictions of both models for each class and selecting the class with the higher average probability. Interestingly, the results from this averaging method mirrored those of the highest probability method, indicating no significant difference in performance between these two approaches.

The third method explored different weighting combinations for audio and text predictions, seeking the optimal blend based on the best average F-Score. The best results were achieved with a weighting of 60\% for audio and 40\% for text. Compared to the first two fusion methods, the weighted approach showed improvements in nearly all metrics. Specifically, it exhibited a higher precision and F-Score for negative valence and slightly higher recall and F-Score for positive valence.

In summary, incorporating both audio and text into the sentiment analysis of music yielded better results than analyzing these modalities separately. Among the fusion methods tested, the weighted approach with a 60\% audio and 40\% text ratio emerged as the most effective strategy. While text models slightly outperformed audio models and fusion methods in classifying negative emotions, the combined audio-text results notably excelled in identifying positive emotions. However, we would also like to note that many of the most interesting cases occur when audio and lyrics exhibit contrasting emotions. This is often artistically intentional. Our approach can also serve to discover such cases.

\section*{Conclusion \& future work}
\label{sec:conclusion}

In this paper, we investigated the sentiment analysis of text and audio in music, developing an approach to combine the results from both modalities. While individual model experiments show acceptable performance, we found weaknesses in classifying negative emotions. We then tested simple methods for combining audio and text results, most effectively with a 60\% audio and 40\% text weighting, emphasizing the importance of considering both modalities. The analysis of misclassifications and discrepancies between audio and text classifications revealed potential errors in annotation adjustments. The significant influence of differences in emotion taxonomy and necessary adjustments likely affected the results.

Challenges in MER and study limitations include the lack of a standardized emotion taxonomy, subjective perception of music, and limited research focusing on simultaneous analysis of text and audio for the same music pieces. This work lays a foundation for future research and development of new approaches in this area.

Music sentiment analysis presents a complex field with significant potential despite numerous challenges. Future research could refine models by training them to classify songs into the four quadrants of Russell's circumplex model and enhance the approach for combining modalities by developing a model that recognizes emotions conveyed in music through both lyrics and audio. However, the quality of data used for model training is crucial. Creating a high-quality, comprehensive dataset could be considered, as existing datasets are often small and focused on either text or audio, with ambiguity regarding which modality existing annotations refer to and neglect of potential contradictions between text and audio emotions. A potential solution could be the combination of existing high-quality datasets created separately for text and audio, as tested on a small scale in this paper. This step is vital given the scarcity of large, high-quality bimodal datasets to advance bimodal sentiment analysis of music.

In addition to refining existing models and developing new approaches for combining audio and text analyses, a promising direction for future research involves the design and training of novel multimodal models capable of processing text and audio inputs both individually and jointly. This approach could potentially offer a more nuanced understanding of the sentiments expressed in music, exploring how different modalities complement or contrast with each other in conveying emotional expressions. Moreover, a multimodal model that integrates text and audio could overcome some of the limitations associated with analyzing these modalities separately, such as discrepancies in sentiment classification between lyrics and melody. By leveraging advancements in multimodal deep learning, researchers could develop systems that better capture the complex interplay of musical elements and lyrical content in expressing sentiment, opening up new possibilities for music emotion recognition and analysis.

\let\oldthebibliography\thebibliography
\let\endoldthebibliography\endthebibliography
\renewenvironment{thebibliography}[1]{
  \begin{oldthebibliography}{#1}
    \setlength{\itemsep}{0em}
    \setlength{\parskip}{0em}
}
{
  \end{oldthebibliography}
}

\bibliographystyle{abbrv}
\bibliography{references}
\end{document}